\documentstyle[twoside,fleqn,npb,epsfig]{article}
\newcommand{\Li}{{\rm Li}}
\newcommand{\Sf}{{\rm S}_{1,2}}
\newcommand{\ds}{\displaystyle}

\newcommand{\be}{\begin{equation}}
\newcommand{\ee}{\end{equation}}
\newcommand{\ba}{\begin{eqnarray}}
\newcommand{\ea}{\end{eqnarray}}
\newcommand{\baz}{\begin{eqnarray*}}
\newcommand{\eaz}{\end{eqnarray*}}

\title{Harmonic Sums and Mellin Transforms}

\author{Johannes Bl\"umlein\address{ 
DESY Zeuthen, D-15738 Zeuthen, Germany}}

\begin{document}

\begin{abstract}
The finite and infinite harmonic sums form the general basis for the
Mellin transforms of all individual functions $f_i(x)$ describing 
inclusive quantities such as coefficient and splitting functions which
emerge in massless field theories. We discuss the mathematical structure
of these quantities.
\end{abstract}

% typeset front matter (including abstract)
\maketitle

%%%%%%%%%%%%%%%%%%%%%%%%%%%%%%%%%%%%%%%%%%%%%%%%%%%%%%%%%%%%%%%%%%%%%%%%%
\section{INTRODUCTION}
%%%%%%%%%%%%%%%%%%%%%%%%%%%%%%%%%%%%%%%%%%%%%%%%%%%%%%%%%%%%%%%%%%%%%%%%%

\noindent
The splitting and coefficient functions in massless QED and QCD can be
evaluated in terms of Nielsen-integrals~\cite{NIEL}
%------------------------------------------------------------------------
\begin{eqnarray}
\lefteqn{S_{n,p}(x) =}
\nonumber\\ & &
\frac{(-1)^{n+p-1}}{(n-1)!p!} \int_0^1 \frac{dz}{z}
\log^{n-1}(z) \log^p(1-zx)
\end{eqnarray}
%------------------------------------------------------------------------
and their Mellin convolutions
%------------------------------------------------------------------------
\begin{eqnarray}
\lefteqn{
f_1(x) \otimes f_2(x) =}\nonumber\\  & &
 \int_0^1 dx_1 \int_0^1 dx_2 \delta(x-x_1 x_2)
f_1(x_1) f_2(x_2)
\end{eqnarray}
%------------------------------------------------------------------------
up to the level of the second order in the coupling constant. 
Alternatively to the $x$--space representation one may consider the
Mellin transform~\cite{MEL}
of these expressions
%------------------------------------------------------------------------
\begin{eqnarray}
{\rm\bf M}[f(x)](N) = \int_0^1 dx x^{N-1} f(x)~.
\end{eqnarray}
%------------------------------------------------------------------------
Here $N$ refers either to the even or odd positive integers depending on
the quantity being studied. The Mellin transforms can be expressed
in terms of polynomials of finite multiple (alternating or
non--alternating) harmonic sums
%------------------------------------------------------------------------
\begin{eqnarray}
S_{k_1,...,k_m}(N) =  \prod_{i=1}^{m} \sum_{n_i = 1}^{n_{i-1}}\frac{
({\rm sign}(k_i))^{n_i}}{{n_i}^{|k_i|}},
\end{eqnarray}
%------------------------------------------------------------------------
with $n_0 \equiv N.$
The finite harmonic sums form classes which are given by their
transcendentality $t = \sum_{l=1}^m |k_l|$. This term was chosen to
characterize the behavior of these sums for $N \rightarrow \infty$.
In this case the values of the sums, if existing, are described by
a transcendental number of rank $t$, i.e. $\log 2, \zeta(2), \zeta(3)$
or $\log 2 \cdot \zeta(2)$, etc. for $t = 1,2,3, ...$ All harmonic sums
have a linear representation in terms of polynomials of harmonic sums
of lower transcendentality and a Mellin transform of a higher function,
which is up to two--loop order always representable as a product
of Nielsen integrals with a polynomial argument structure. Moreover
algebraic relations between harmonic sums of a given transcendentality
allow for a considerable reduction of the basic set of functions which 
has to be calculated and leads to further structural simplifications 
as well.
\vspace*{-3mm}
%%%%%%%%%%%%%%%%%%%%%%%%%%%%%%%%%%%%%%%%%%%%%%%%%%%%%%%%%%%%%%%%%%%%%%%%%
\section{LINEAR REPRESENTATIONS}
%%%%%%%%%%%%%%%%%%%%%%%%%%%%%%%%%%%%%%%%%%%%%%%%%%%%%%%%%%%%%%%%%%%%%%%%%
\noindent
The number of alternating and non-alternating harmonic sums of 
transcendentality $k$ is  $n_S(k) = 2 \cdot 3^{k-1}$, yielding a total
number of sums of $3^k - 1$.
The harmonic sums obey  representations in terms of multiple integrals
which are obtained from
%-----------------------------------------------------------------------
\begin{eqnarray}
\label{eqSc}
S_k(N) &=& \int_0^1 dx \frac{\left[-\log(x)\right]^{k-1}}
{(k-1)!}
\frac{x^N-1}{x-1}
\\
\label{eqSd}
S_{-k}(N) &=&  \int_0^1 dx \frac{\left[-\log(x)\right]^{k-1}}
{(k-1)!}
\frac{(-x)^N-1}{x+1}
\end{eqnarray}
%------------------------------------------------------------------------
consecutively. These integrals are     expressed by harmonic sums
of lower degree and a Mellin transform which is not further reducible
on the linear level, cf.~\cite{BK1} for details. In this representation
also infinite harmonic sums occur, which can be represented as linear
combinations of basic transcendentals as $\log(2), \zeta(2), \zeta(3),
\Li_4(1/2)$ up to the level of transcendentality 4. These sums were given 
in explicit form in Ref.~\cite{BK1}.
Already up to transcendentality 4 Mellin transforms of rather
complicated functions, as e.g.
%------------------------------------------------------------------------
\begin{eqnarray}
\label{eqF1}
\lefteqn{F_1(x) = \Sf\left(\frac{1-x}{2}\right) + \Sf(1-x)} \\
& &  - \Sf\left(\frac{1-x}{1+x}\right) + \Sf\left(\frac{1}{1+x}\right) 
\nonumber\\ & &
- \log(2)
\Li_2\left(\frac{1-x}{2}\right) \nonumber\\  & &
+ \frac{1}{2} \log^2(2) \log\left(
\frac{1+x}{2}\right) 
- \log(2) \Li_2\left(\frac{1-x}{1+x}\right)  
\nonumber
\end{eqnarray}
%---------------------------------------------------------------------
emerge.
These Mellin transforms, however, turn out to be reducible using the
algebraic relations given below, i.e. $F_1(x)$ can be obtained as 
a linear combination of
Mellin convolutions of more elementary functions and simpler argument
structure. This highlights the importance of algebraic relations between
the harmonic sums, to which we are turning now. 
\vspace*{-3mm}
%%%%%%%%%%%%%%%%%%%%%%%%%%%%%%%%%%%%%%%%%%%%%%%%%%%%%%%%%%%%%%%%%%%%%%%%%
\section{ALGEBRAIC RELATIONS}
%%%%%%%%%%%%%%%%%%%%%%%%%%%%%%%%%%%%%%%%%%%%%%%%%%%%%%%%%%%%%%%%%%%%%%%%%
\noindent
The finite harmonic sums of order $k$ are related by algebraic
equations. They are obtained studying sums of harmonic sums with
permutations in the set of their indices. The simplest relation is due
to {\sc Euler}~\cite{EUL} for two indices
%------------------------------------------------------------------------
\begin{equation}
S_{m,n} + S_{n,m} = S_m S_n + S_{m \wedge n},
\end{equation}
%------------------------------------------------------------------------
where 
$m_1 \wedge m_2 \wedge ... m_k~=~\prod_{l=1}^k~
{\rm sign}(m_k)~\sum_{l=1}^k |m_k|$. The corresponding relations for 
three- and fourfold
sums are~\cite{BK1}
%------------------------------------------------------------------------
\begin{equation}
\label{threef}
\sum_{\rm perm} S_{l,m,n} = S_l S_m S_n 
+ \sum_{\rm perm} S_l S_{m \wedge n}
+ S_{l \wedge m \wedge n}
\end{equation}
%------------------------------------------------------------------------
and
%------------------------------------------------------------------------
\begin{eqnarray}
\sum_{\rm perm} S_{k,l,m,n} &=& S_k S_l S_m S_n
+ \sum_{\rm perm} S_k S_l S_{m \wedge n} \nonumber\\
&+& \sum_{\rm perm} S_{k \wedge l} S_{m \wedge n} \\
&+& 2 \sum_{\rm perm} S_k S_{l \wedge m \wedge n} 
+ 6 S_{k \wedge l \wedge m \wedge n}. \nonumber
\end{eqnarray}
%------------------------------------------------------------------------
For non-alternating sums and $N \rightarrow \infty$  relation 
(\ref{threef})
was given in \cite{IND} before.

Harmonic sums with the same index, but arbitrary length of the
index set
are related
to  determinant--structures, cf.~\cite{BK1}, and can be  calculated
in terms of products of single harmonic sums only, as e.g.
%------------------------------------------------------------------------
\begin{eqnarray}
S_{\underbrace{\mbox{\scriptsize -1, \ldots ,-1}}_{\mbox{\scriptsize
$k$}}} &=& \frac{1}{k} \sum_{l=1}^k S_{(-1)^l |l|} 
S_{\underbrace{\mbox{\scriptsize -1, \ldots ,-1}}_{\mbox{\scriptsize
$k-l$}}} \\
S_{\underbrace{\mbox{\scriptsize 1, \ldots ,1}}_{\mbox{\scriptsize
$k$}}} &=& \frac{1}{k} \sum_{l=1}^k S_{l} 
S_{\underbrace{\mbox{\scriptsize 1, \ldots ,1}}_{\mbox{\scriptsize
$k-l$}}}.
\end{eqnarray}
%---------------------------------------------------------------------
Starting with 3-fold harmonic sums more algebraic relations can be
obtained by partial permutations of the index set. For 3-fold 
alternating or non-alternating harmonic sums 3 relations are 
obtained~\cite{BG,BK1}, which cover the case (\ref{threef})
and result
from the combinations of
%------------------------------------------------------------------------
\begin{eqnarray}
\label{Ta}
T &=& S_{a,b,c} + S_{a,c,b} - S_{a \wedge b,c} - S_{a \wedge c,b}
     -S_{a,b \wedge c} \nonumber \\ &+&
      S_{a \wedge b \wedge c} \\
T &=& S_c S_{a,b} - S_{c,a,b} + S_{c,a \wedge b} - S_c S_{a \wedge b} \\
T &=& S_b S_{a,c} - S_{b,a,c} + S_{b,a \wedge c} - S_b S_{a \wedge c} \\
\label{Td}
T &=& S_{b,c,a} + S_{c,b,a} - S_{b \wedge c, a}  - S_c S_{b,a}
     + S_b S_{a,c}  \nonumber\\ &-&
       S_b S_{a \wedge c}.
\end{eqnarray}
%------------------------------------------------------------------------
Using these relations the number of Mellin transforms occurring in the
linear representations can be reduced substantially. Moreover Mellin
transforms of more complicated functional structure are recognized as
transforms of convolutions of much more elementary functions. Up to
2--loop order only simple, reducible variants of harmonic sums of the
type $S_{\pm 1,\pm 1,\pm 1, \pm 1}(N)$ occur. The remaining linear 
representations can be represented by the Mellin-transforms of the
basic
functions given below.

For the analytic continuation of the Mellin moments {\bf M}$[f_i(x)](N)$
in addition to the
              well--known relations for single harmonic sums only
 the          Mellin transforms of 24 basic functions
have to be analytically continued, see
\cite{JB1}.

In Ref.~\cite{BK1} a systematic evaluation of the Mellin transforms of
the individual functions representing the polarized and unpolarized
coefficient functions and
anomalous dimensions up to 2--loop order (cf. e.g.~\cite{NZ}) are given.
They can be expressed through harmonic sums recursively, which are
linear functions of the Mellin transforms of the functions listed below.
About 80 functions $f_i(x)$ occur. One example is
%------------------------------------------------------------------------
{\small
\begin{eqnarray}
\lefteqn{
{\rm\bf M}\left[
{\ds \frac{1}{1+z}\Biggl[ \Li_{3}\left(\frac{1-z}{1+z}\right)}
{\ds -\Li_{3}\left(-\frac{1-z}{1+z}\right)}\Biggr]\right](N)
 =}
\nonumber\\ & &
{\ds (-1)^{N-1} \biggl\{
 S_{1,1,-2}(N-1) - S_{1,-1,2}(N-1)}
  \nonumber\\ & &
{\ds   + S_{-1,1,2}(N-1)  - S_{-1,-1,-2}(N-1) } \nonumber\\  & &
{\ds
+ 2\zeta(2)S_{1,-1}(N-1)
 + \frac{1}{4}\zeta(2)S_{1}^{2}(N-1)}   \nonumber\\
& &  {\ds - \frac{1}{4}\zeta(2)S_{-1}^{2}(N-1)
 - \zeta(2)S_{1}(N-1)S_{-1}(N-1)}
 \nonumber\\ & & {\ds
 - \zeta(2)S_{-2}(N-1)   
 -\left[ \frac{7}{8}\zeta(3) - \frac{3}{2}\zeta(2)\log
  2\right]} \nonumber \\ & & \times S_{1}(N-1)
{\ds +\left[ \frac{21}{8}\zeta(3) - \frac{3}{2}\zeta(2)\log
  2\right] S_{-1}(N-1)}   \nonumber\\
& &  {\ds -2\Li_{4}\left(\frac{1}{2}\right)
 + \frac{19}{40}\zeta^{2}(2) + \frac{1}{2}\zeta(2)\log^{2}2
 - \frac{1}{12}\log^{4}2 \biggr\} }   \nonumber
\end{eqnarray}
}
%-----------------------------------------------------------------------
\vspace*{-0.7cm}
%%%%%%%%%%%%%%%%%%%%%%%%%%%%%%%%%%%%%%%%%%%%%%%%%%%%%%%%%%%%%%%%%%%%%%%%%
\section{CONCLUSIONS}

%\vspace{-8cm}
%%%%%%%%%%%%%%%%%%%%%%%%%%%%%%%%%%%%%%%%%%%%%%%%%%%%%%%%%%%%%%%%%%%%%%%%%

\noindent
A systematic study of the finite harmonic alternating and 
non--alternating sums has been performed up to the four--fold sums,
which have been evaluated in explicit form in the linear representation.
Algebraic relations were used to reduce this set to a representation
over a much smaller  set of functions. Whereas in the $N$--space 
representations these relations are truly algebraic, the corresponding
relations in the $x$--space representation are given by sums of 
multiple Mellin convolutions. In this representation the algebraic
relations lead to essential structural simplifications both concerning
the contributing functions as well as their argument structure. The 
corresponding
Mellin transforms were evaluated in explicit  form.

\noindent
{\bf Acknowledgment.}
The work was supported         by EU contract FMRX-CT98-0194(DG 12 -
MIHT).
\nopagebreak
%%%%%%%%%%%%%%%%%%%%%%%%%%%%%%%%%%%%%%%%%%%%%%%%%%%%%%%%%%%%%%%%%%%%%%%%

%%%%%%%%%%%%%%%%%%%%%%%%%%%%%%%%%%%%%%%%%%%%%%%%%%%%%%%%%%%%%%%%%%%%%%

\noindent
%------------------------------------------------------------------------
\begin{eqnarray}
%------------------------------------------------------------------------
{\small
\begin{array}{ll}
{\ds \frac{\log(1+x)}{x+1}} & {\ds \frac{\log^2(1+x)-\log^2(2)}{x-1}} \\
&   \\
%------------------------------------------------------------------------
{\ds \frac{\log^2(1+x)}{x+1}} & {\ds \frac{\Li_2(x)}{x+1}}\\
&  \\
%------------------------------------------------------------------------
{\ds \frac{\Li_2(x)-\zeta(2)}{x-1}} &
{\ds \frac{\Li_2(-x)}{x+1}} \\ 
&  \\
%------------------------------------------------------------------------
{\ds \frac{\Li_2(-x)+\zeta(2)/2}{x-1}} &
{\ds \frac{\log(x)\Li_2(x)}{x+1}} \\
&  \\
%------------------------------------------------------------------------
{\ds \frac{\log(x)\Li_2(x)}{x-1}} & {\ds \frac{\Li_3(x)}{x+1}} \\ 
& \\
%------------------------------------------------------------------------
{\ds \frac{\Li_3(x)-\zeta(3)}{x-1}} & {\ds \frac{\Li_3(-x)}{x+1}} \\
&  \\
%------------------------------------------------------------------------
{\ds \frac{\Li_3(-x)-3 \zeta(3)/4}{x-1}} &
{\ds \frac{\Sf(x)}{x+1}} \\
&  \\
%------------------------------------------------------------------------
{\ds \frac{\Sf(x)-\zeta(3)}{x-1}} & {\ds \frac{\Sf(-x)-\zeta(3)/8}{x-1}} 
\\
&  \\
%------------------------------------------------------------------------
{\ds \frac{\Sf(-x)}{x+1}} & {\ds \frac{\Sf(x^2)}{x+1}} \\
& \\
%------------------------------------------------------------------------
{\ds \frac{\Sf(x^2)-\zeta(3)}{x-1}} & 
{\ds \log(1-x) \frac{\Li_2(-x)}{x+1}} \\
&  \\
%------------------------------------------------------------------------
{\ds \frac{\log(1+x) - \log(2)}{x-1} \Li_2(-x)} &
{\ds \frac{\log(1-x) \Li_2(x)}{1+x}} \\
\\
&  \\
%------------------------------------------------------------------------
{\ds \frac{\log(1+x) - \log(2)}{x-1} \Li_2(x)} &
{\ds \frac{\log(1+x)}{1+x} \Li_2(x)}\\
&  \\
%------------------------------------------------------------------------
\end{array}
}
\nonumber
\end{eqnarray}
%------------------------------------------------------------------------
\begin{center}
\vspace*{-3mm}
{\sf\small List~of~the~basic~functions}
\end{center}
%%%%%%%%%%%%%%%%%%%%%%%%%%%%%%%%%%%%%%%%%%%%%%%%%%%%%%%%%%%%%%%%%%%%%%%%
\end{document}